\documentclass[12pt]{article}
\usepackage{graphicx}
\usepackage{psfrag}

\begin{document}

\begin{titlepage}
\null\vspace{-62pt}

\pagestyle{empty}
\begin{center}

\vspace{1.0truein} {\Large\bf Symmetry breaking and restoration in Lifshitz type theories}

\vspace{1in}
{\large K.~Farakos and D.~Metaxas} \\
\vskip .4in
{\it Department of Physics,\\
National Technical University of Athens,\\
Zografou Campus, 15780 Athens, Greece\\
kfarakos@central.ntua.gr, metaxas@central.ntua.gr}\\

\vspace{0.5in}

\vspace{.5in}
\centerline{\bf Abstract}

\baselineskip 18pt
\end{center}

\noindent
We consider the one-loop effective potential at zero and finite temperature in scalar field theories with anisotropic space-time scaling. For $z=2$, there is a symmetry breaking term induced at one loop at zero temperature and we find symmetry restoration through a first-order phase transition at high temperature. For $z=3$, we considered at first the case with a positive mass term at tree level and found no symmetry breaking effects induced at one-loop, and then
we study the case with
a negative mass term at tree level where we cannot conclude about
symmetry restoration effects at high temperature because of the
imaginary parts that appear in the effective potential for small values of the scalar field.

\end{titlepage}

\newpage
\pagestyle{plain}
\setcounter{page}{1}
\newpage

\section{Introduction}

Non-relativistic field theories in the Lifshitz context, with anisotropic scaling between temporal and spatial directions, measured by the dynamical critical exponent, $z$,
\begin{equation}
t\rightarrow b^z t,\,\,\,x_i\rightarrow b x_i,
\end{equation}
have been considered recently since they have an improved ultraviolet behavior and their renormalizability properties are quite different than conventional Lorentz symmetric theories \cite{visser}--\cite{iengo}.
Various field theoretical models and extensions of gauge field theories at the Lifshitz point have already been considered
\cite{hor3}.

When extended in curved space-time, these considerations may provide a renormalizable candidate theory of gravity \cite{hor1} and applications of these concepts in the gravitational and cosmological context
have also been widely investigated
\cite{kk}.

We will consider here the case of a single scalar field in flat space-time.
The weighted in the units of spatial momenta scaling dimensions are $[t]=-z$ and $[x_i]=-1$, with $z$ the anisotropic scaling, and $i=1,...,D$ the spatial index (here we consider $D=3$). The action with a single scalar field is
\begin{equation}
S=\int dt d^Dx \left[ \frac{1}{2} \dot{\phi}^2 -\frac{1}{2}\phi(-\Delta)^z \phi-U_0(\phi)\right],
\label{gen}
\end{equation}
with $\Delta=\partial_i^2$ and $[\phi]=\frac{D-z}{2}$.

In order to investigate the various implications of a field theory, in particle physics and cosmology, it is particularly important to examine its symmetry structure, both at the classical and the quantum level, at zero and finite temperature, via the effective action and effective potential \cite{col}-- \cite{dolan}.
We should note that, in order to get information on possible instabilities of the theory, we study the one-loop, perturbative effective potential, given by the one-particle irreducible diagrams of the theory, and not the full, non-perturbative, convex effective potential given by the so-called Maxwell construction \cite{wett2}.

In a recent work \cite{kim1}, the effective potential for a scalar theory was considered for the case of $z=2$ and it was shown that, at one loop order, there is a symmetry breaking term induced quantum mechanically; also the finite temperature effective potential was studied at one loop, and it was argued that there is no symmetry restoration at high temperature.

We study the theory with $z=2$ in Sec.~2 and find at zero temperature a symmetry breaking term at one loop that agrees with the results of \cite{kim1}.
However, we have also studied the finite temperature effective potential both analytically and numerically, and have found the interesting result of symmetry restoration at high temperature through a first-order phase transition. In view of the importance of symmetry breaking phenomena throughout field theory and cosmology, we have also studied the situation for the case of $z=3$ in Sec.~3: in the case of a positive or zero mass term in the tree level we found no symmetry breaking terms induced at one loop. In the case of a negative mass term at the tree level we calculated the full effective potential at high temperature and
found no symmetry restoration effects  induced at one 
loop because of the imaginary parts that appear in the effective potential for small values of the scalar field.

\section{Effective potential for $z=2$ at zero and finite temperature}

We consider the action (\ref{gen}) with $z=2$,
\begin{equation}
S=\int dt d^3x \left[ \frac{1}{2} \dot{\phi}^2 -\frac{1}{2}(\partial_i^2\phi)^2-U_0(\phi)\right].
\end{equation}
Here we have $[\phi]=1/2$ and $U_0(\phi)$, the tree-level potential, is a polynomial up to the weighted marginal power of $\phi$ (here the tenth).
The one-loop contribution to the effective potential,
\begin{equation}
U_1=\frac{1}{2}\int\frac{d^4k}{(2\pi)^4}\ln (k_0^2+k_i^4+U_0'')
      =\frac{1}{4\pi^2}\int k^2 dk \sqrt{k^4+U_0''}
\end{equation}
(where, in the last equation, $k^2=k_i^2$) can be evaluated with a cutoff $\Lambda$ in the spatial momentum via differentiation with respect to $y=U_0''$ (primes denote differentiation with respect to $\phi$).
We get
\begin{equation}
\frac{d^2 U_1}{dy^2}=-\frac{1}{16\pi^2}\frac{1}{y^{3/4}}\int_0^{\infty}dx\frac{x^2}{(x^4+1)^{3/2}}
\end{equation}
and, using the boundary conditions
\begin{equation}
\frac{d U_1}{dy}(y=0)=\frac{\Lambda}{8\pi^2},\,\,\,U_1(y=0)=\frac{\Lambda^5}{20\pi^2},
\end{equation}
we get
\begin{equation}
U_1(\phi)=\frac{1}{8\pi^2}U_0'' \,\Lambda \, - \, c (U_0'')^{5/4},
\end{equation}
where $c= \frac{1}{4\pi^2} \int_0^{\infty}dx\frac{x^2}{(x^4+1)^{3/2}} =\Gamma(3/4)^2/10\pi^{5/2}$.
The first term, which is linearly divergent, can be renormalized with appropriate counterterms in the potential, and the second term, which is generally negative, can lead to a non-zero minimum, even if the original potential had a unique minimum at $\phi=0$.

We consider here the case of a massless theory, with a single relevant operator, $U_0(\phi)=\frac{\lambda}{4!}\phi^4$,
and add the counterterms $\frac{1}{2}A\phi^2+\frac{1}{4!}B\phi^4$. The condition $U''(0)=0$ eliminates the quadratic terms and, because of the infrared divergence, the condition at a non-zero $\phi=\alpha$, $U''''(\alpha)=\lambda$, has been imposed. Since $[\lambda]=3$ and $[\phi]=1/2$, we write $\alpha^2 =\mu$ and $\lambda=\tilde{\lambda}\mu^3$, in terms of an overall mass scale $\mu$.

The full effective potential after renormalization is
\begin{equation}
U(\phi)=\frac{\lambda}{4!}\left(1-\frac{15c\tilde{\lambda}^{1/4}}{2^5\cdot2^{1/4}}\right)\phi^4 \,-\,c \left(\frac{\lambda}{2}\phi^2\right)^{5/4}.
\label{res1}
\end{equation}

The full effective potential now has a non-zero minimum, and a mass term will be generated after expansion around this minimum, but it should be noted that the situation is not entirely analogous to the usual Coleman-Weinberg mechanism, since the tree-level potential has a dimensionful parameter already. The situation is similar if other relevant operators with dimensionful couplings are considered ($\phi^6$ and $\phi^8$) but not if only the marginal $\phi^{10}$ operator, with a dimensionless coupling is considered in the tree-level potential. These results agree with the corresponding conclusions from \cite{kim1}. We now proceed to the calculation of the finite temperature effects and show that when the appropriate corrections to the effective potential are taken into account, there appear to exist symmetry restoration effects at high temperature, and indeed with a first-order phase transition.

The one-loop effective potential at finite temperature is \cite{dolan}
\begin{equation}
U_{1T}=\frac{1}{2\beta}\sum_n \int\frac{d^3k}{(2\pi)^3}\ln
\left(\frac{4\pi^2n^2}{\beta^2}+E^2\right),
\end{equation}
where $\beta = 1/T$ is the inverse temperature, $E^2=k^4+U_0''$ and the sum is over non-negative integers, $n$.
Using
\begin{equation}
\sum_n\ln\left(\frac{4\pi^2n^2}{\beta^2}+E^2\right)=
2\beta\left[\frac{E}{2}+\frac{1}{\beta}\ln(1-e^{-\beta E})\right],
\end{equation}
the total effective potential can be written as
$U_{1T}=U_1 + U_T$, where $U_1$ is the zero temperature contribution analyzed before and the temperature-dependent part is
\begin{equation}
U_T(\phi)=T \int\frac{d^3 k}{(2\pi)^3} \ln \left( 1- e^{-\beta\sqrt{k^4+U_0''}} \right)
\end{equation}
and is manifestly real for all values of $\phi$.
The total effective potential can be plotted for various temperatures (at fixed $\lambda$) and the results indicate symmetry restoration at high temperature with a first-order phase transition.

As an example, the full potential is plotted in Fig.~1 for various temperatures. All quantities are rescaled in terms of appropriate powers of the dimensionful constant $\mu$ introduced before: the potential in units of $\mu^5$, the temperature in units of $\mu^2$ and $\phi$ in units of
$\mu^{1/2}$. The results are shown
for temperatures $\tilde{T}=0.008, 0.01, 0.012$ where $T=\tilde{T}\mu^2$
(we took $\tilde{\lambda}=0.1$). Also a constant, temperature-dependent term discussed below is not shown.

\vspace{0cm}

\begin{figure}[!t]
\begin{center}
\includegraphics[scale=1]{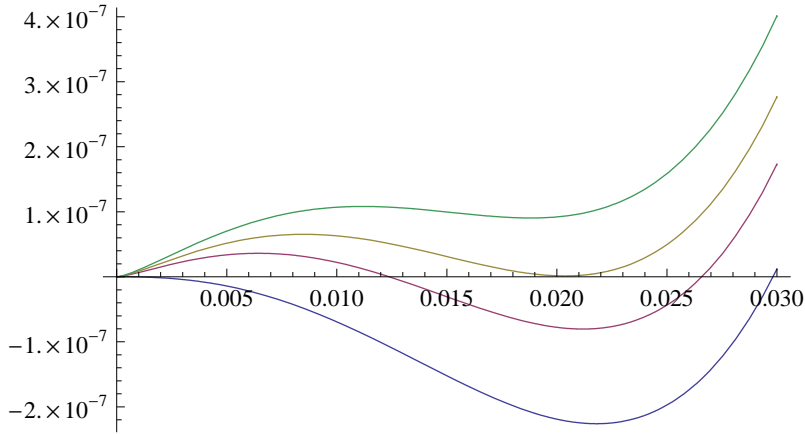}
\caption{The exact expression for the effective potential as a function of $\phi$
 for the theory with $z=2$,
at one loop at zero and  finite temperature.
The potential is plotted in units of $\mu^5$ and $\phi$ in units of
$\mu^{1/2}$.The temperature is in units of $\mu^2$,
the lowest curve is the potential at zero temperature and the other three curves are at increasing temperatures $\tilde{T}=0.008, 0.01, 0.012$ where $T=\tilde{T}\mu^2$
(we took $\tilde{\lambda}=0.1$).}
\end{center}
\end{figure}

In \cite{kim1}  it was argued that no such phase transition takes place. However, the conclusions of \cite{kim1} were based on the examination of the second derivative of the potential at the origin. This is not an appropriate test at this case because of the non-analytical terms that appear in the potential at the finite-temperature limit.
One can see the possible such terms that may arise by considering an analytical approximation with an infrared momentum cutoff for the effective potential at finite temperature:
in order to obtain this approximation, we can expand the logarithm in the previous equation and write
\begin{equation}
U_T(\phi)=-\frac{T^{5/2}}{2\pi^2}\int dx\,x^2 \sum_n \frac{1}{n} e^{-n\sqrt{x^4+\beta^2 U_0''}}.
\label{app1}
\end{equation}
Now, in the exponent in the previous expression, it turns out to be a good approximation to use
$(x^4 +a^2)^{1/2}\approx x^2 + \frac{a^2}{2 x^2}$ (where $a^2 =\beta^2 U_0''$).
The integrand in (\ref{app1}) is maximum for values of  $x\approx 1$ (for small values of $n$) so our approximation to consider the main contribution for $x^2>a$ is valid since we are interested in the high-temperature regime $a<<1$.
It is definitely a good approximation for large $x$, and, since for small $x$ the exponential goes to zero, it is equivalent to an  effective infrared momentum cutoff.
Then the sum can be done with the help of the elementary integral
$\int dx e^{-c_1 x^2- \frac{c_2}{x^2}}=\frac{1}{2}\sqrt{\frac{\pi}{c_1}}e^{-2\sqrt{c_1 c_2}} $.
We get
\begin{equation}
U_T=-\frac{T^{5/2}}{8\pi^{3/2}}\sum_n\left(\frac{1}{n^{5/2}}
+\frac{\sqrt{2}a}{n^{3/2}}\right)
e^{-\sqrt{2} an}.
\end{equation}
The sums can be expressed in terms of the polylog function
$P_\nu(w)=\sum_n \frac{1}{n^\nu} w^n$, with $w=e^{-\sqrt{2} a}$. In the high-temperature regime, where $a<<1$, one can expand around $w=1$ using the Taylor expansions
\begin{equation}
P_{5/2}(w)=\zeta(5/2) + \zeta(3/2)(w-1) +\frac{4}{3}i\sqrt{\pi}(w-1)^{3/2}+ O(w-1)^2,
\end{equation}
\begin{equation}
P_{3/2}(w)=\zeta(3/2) +2i\sqrt{\pi}(w-1)^{1/2}+ \zeta(1/2)(w-1) + O(w-1)^{3/2},
\end{equation}
The leading terms of the final result at high temperature are
\begin{equation}
U_T(\phi)= -\frac{\zeta(5/2)}{8 \pi^{3/2}}\, T^{5/2} + \frac{2^{3/4}}{12\pi} T\,(U_0'')^{3/4},
\label{res2}
\end{equation}
which shows, besides a constant term, a second positive term, linear in the temperature, which will dominate, in the high-temperature limit, the previous, symmetry-breaking, zero-temperature terms.
The first, constant term is exact and is not shown in the figures. It corresponds to the black-body radiation term that is proportional to $T^4$ in the usual Lorentz-invariant case with $z=1$.
The second, non-analytical term was not found in \cite{kim1}, where the condition with the second derivative of the potential at the origin was used in order to determine the critical temperature. It is clear from the above form of the potential that, since the extra term is non-analytical, and of the form $\phi^{3/2}$ for the interaction $\lambda \phi^4$, this condition cannot be used reliably in this case.

In Fig.~2 we show the full effective potential in the analytical approximation with the infrared cutoff, with the same parameters as in Fig.~1, and we see a flattening of the potential compared with the exact evaluation, the general features of the phase transition, however, remain the same. In fact, even the full quantitative features of the phase transition may be more appropriately investigated using a coarse-grained potential with an infrared cutoff of the form used here \cite{wett}.
 It is clear from these results that one has the interesting phenomenon of symmetry restoration at high temperature, with a potential term that indicates a first-order phase transition.

\vspace{0cm}

\begin{figure}[ht]
\begin{center}
\includegraphics[scale=1]{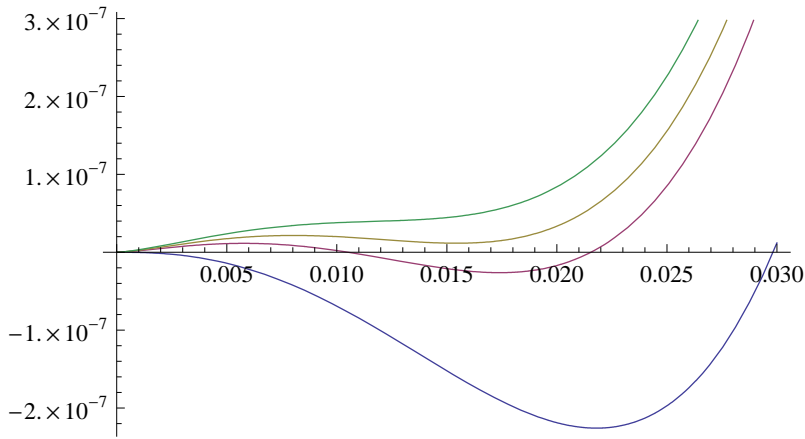}
\caption{Same as Fig.~1, with the same parameters, using the analytical approximation with the infrared cutoff of Eq.~(\ref{res2})}
\end{center}
\end{figure}

\section{Effective potential for $z=3$}

In view of the previous results it would be interesting to investigate similar effects in the case with anisotropic scaling $z=3$; unfortunately we find no indication of symmetry-breaking terms in one-loop order.
The action (\ref{gen}) with $z=3$ is
\begin{equation}
S=\int dt d^3x \left[ \frac{1}{2} \dot{\phi}^2 -\frac{1}{2}(\partial_i \nabla^2\phi)^2-U_0(\phi)\right]
\end{equation}
where, now, $[\phi]=0$ and the potential, $U_0$, can be an arbitrary function of $\phi$. Taking, for definitiveness,
\begin{equation}
U_0(\phi) =\frac{1}{2} m^2 \phi^2 + \frac{1}{4!}\lambda\phi^4,
\end{equation}
with $[\lambda]=[m^2]=6$, one gets for the one-loop contribution to the effective potential, with the momentum cut-off, $\Lambda$,
\begin{equation}
U_1(\phi) = \frac{1}{12 \pi^2} \left[ \frac{\Lambda\sqrt{\Lambda^2+ U_0''}}{2}
      +\frac{U_0''}{2}\ln (\Lambda +\sqrt{\Lambda^2+U_0''})
      -\frac{U_0''}{2}\ln (\sqrt{U_0''}) \right],
\end{equation}
which, for large $\Lambda$, becomes
\begin{equation}
U_1(\phi)=\frac{1}{24\pi^2}\left[ U_0'' \ln(2\Lambda)-\frac{U_0''}{2}\ln U_0''\right].
\end{equation}

In the case of $m^2>0$, where there is no symmetry breaking at tree level, we add the counterterms
$\frac{1}{2}A\phi^2+\frac{1}{4!}B\phi^4$ and impose the renormalization conditions $U''(0)=m^2$ and
$U''''(0)=\lambda$, to get
\begin{equation}
U(\phi) = \frac{1}{2} m^2\left(1+\frac{1}{48\pi^2}\frac{\lambda}{m^2}\right) \phi^2 +
        \frac{\lambda}{4!}(1+\frac{3}{48\pi^2}\frac{\lambda}{m^2}) \phi^4
        -\frac{1}{48\pi^2} U_0''(\phi)
        \ln\frac{U_0''(\phi)}{m^2}.
\label{res3}
\end{equation}
One may take the $m^2\rightarrow 0$ limit in this expression, remembering that the dimensionless coupling constant, which is $\lambda / m^2$, is to be kept fixed and small. Doing that, one can easily see that the resulting potential is everywhere positive, without signs of any symmetry-breaking effects (the same conclusion holds for all positive values of $m^2$).

In the case of $m^2<0$, where there is symmetry breaking at tree level, after adding the same counterterms,
we can take renormalization conditions at the minimum $\phi=\sigma$, with $\sigma^2=-6m^2/\lambda$.
Imposing the conditions $U'(\sigma)=0$, $U''(\sigma)=-2m^2$, which preserve the tree level mass and minimum,
we get
\begin{equation}
U(\phi) = \frac{1}{2} m^2\left(1-\frac{1}{96\pi^2}\frac{\lambda}{m^2}\right) \phi^2 +
        \frac{\lambda}{4!}(1-\frac{1}{32\pi^2}\frac{\lambda}{m^2}) \phi^4
        -\frac{1}{48\pi^2} U_0''(\phi)
        \ln\frac{U_0''(\phi)}{U_0''(\sigma)}.
\label{res4}
\end{equation}

Now the finite temperature effective potential,
\begin{equation}
U_T(\phi)=T \int\frac{d^3 k}{(2\pi)^3} \ln \left( 1- e^{-\beta\sqrt{k^6+U_0''}} \right),
\end{equation}
can be calculated via
\begin{equation}
\frac{\partial U_T}{\partial a^2} =\frac{T^2}{12 \pi^2}\int dx
\frac{1}{\sqrt{x^2+a^2}}\frac{1}{e^{\sqrt{x^2+a^2}}-1},
\end{equation}
where $a^2=\beta^2 U_0''$, and expanded
in the high-temperature limit using formulas from \cite{dolan}; the temperature-dependent part of the potential is
\begin{equation}
U_T(\phi)=-\frac{T^2}{36}+\frac{T}{12\pi} \sqrt{U_0''(\phi)} +
\frac{U_0''(\phi)}{48\pi^2}\ln \frac{U_0''(\phi)}{c_B T^2}
-\frac{\zeta(3)}{384\pi^4}\frac{U_0''(\phi)^2}{T^2}+\cdots,
\end{equation}
where the first term is the $\phi$-independent black-body radiation term,
$\ln c_B =1-2\gamma+2\ln 4\pi$, $\gamma=0.577...$, and subsequent terms are of higher order in
$U_0''(\phi)/T^2$.
Only the second term in this expansion can lead to symmetry restoration at high temperature,
we see, however, that, for negative $m^2$,
one cannot make any such conclusion because of the imaginary parts that appear in the expression for the effective potential for values of $\phi$ near zero.

\section{Comments}

In this work we studied the symmetry breaking effects of the one-loop effective potential at zero temperature for theories
of the Lifshitz type with a single scalar field
with anisotropic scaling $z=2$ and $z=3$, and the possible symmetry restoration effects at high temperature.

In the case of $z=2$ we found symmetry breaking terms induced at one loop at zero temperature (in agreement with a previous work \cite{kim1}) and we studied the effective potential at finite temperature at one loop, both numerically and analytically through an approximation that is equivalent to imposing an infrared cutoff, and may be useful for future studies of applications of these theories in other field theoretical or cosmological contexts. We found the interesting effects of symmetry restoration at high temperature, through an apparently first-order phase transition.

Because of the importance of symmetry breaking and restoration phenomena in quantum field theory and cosmology we also studied the case of scalar field theory with $z=3$ but found no similar effects: in the case where we have a positive or zero mass term at the potential at tree level we calculated the one-loop contribution to the
effective potential and found no symmetry breaking terms induced at this level. In the case where there is a negative mass term in the potential (with symmetry breaking at tree level) we calculated the full effective potential at high temperature and found that there is no conclusion of symmetry restoration at high temperature because of the imaginary parts that appear in the expression for the effective potential for values of the field near the origin.

In view of the above results it would be interesting to include gauge fields and consider the symmetry breaking and restoration effects including gauge field interactions in a future investigation.

\vspace{0.5in}

\centerline{\large\bf  Acknowledgements} \noindent
We would like to thank Jean Alexandre for several discussions and for bringing into our attention the results of \cite{kim1}.

\end{document}